\documentclass[amsmath,amssymb,aps,pra,floatfix,reprint,superscriptaddress]{revtex4-1}
\usepackage{color}
\usepackage{graphicx}
\usepackage{comment}
\usepackage{multirow}
\usepackage{mathdots}
\usepackage{bm}
\usepackage{subcaption}
\usepackage[normalem]{ulem}

\usepackage{url}

\usepackage[breaklinks=true,hidelinks]{hyperref}

\begin{document}

\title{Control of photodissociation with dynamic Stark effect induced by Thz pulses}

\author{A. T\'oth}\email{Attila.Toth@eli-alps.hu}
\affiliation{ELI-ALPS, ELI-HU Non-Profit Ltd,H-6720 Szeged, Dugonics t\'{e}r 13, Hungary}

\author{A. Csehi}
\affiliation{Department of Theoretical Physics, University of Debrecen, H-4010 Debrecen, PO Box 5, Hungary}
\affiliation{ELI-ALPS, ELI-HU Non-Profit Ltd,H-6720 Szeged, Dugonics t\'{e}r 13, Hungary}

\author{G. J. Hal\'asz}
\affiliation{Department of Information Technology, University of Debrecen, H-4010 Debrecen, PO Box 12, Hungary}

\author{\'A. Vib\'ok }\email{Vibok@phys.unideb.hu}
\affiliation{Department of Theoretical Physics, University of Debrecen, H-4010 Debrecen, PO Box 5, Hungary}
\affiliation{ELI-ALPS, ELI-HU Non-Profit Ltd,H-6720 Szeged, Dugonics t\'{e}r 13, Hungary}



\begin{abstract}
We demonstrate how dynamic Stark control (DSC) can be achieved on molecular 
photodissociation in the dipole limit, using single-cycle (FWHM) laser pulses in the 
terahertz (THz) regime. As the laser-molecule interaction follows the instantaneous 
electric field through the permanent dipoles, the molecular potentials dynamically 
oscillate and so does the crossings between them. In this paper, we consider rotating-vibrating 
diatomic molecules (2D description) and reveal the interplay between the dissociating 
wave packet and the dynamically fluctuating crossing seam located in the configuration 
space of the molecules spanned by the R vibrational and $\theta$ rotational coordinates.
Our showcase example is the widely studied lithium-fluoride (LiF) molecule for 
which the two lowest $\Sigma$ states are nonadiabatically coupled at an avoided 
crossing (AC), furthermore a low-lying pure repulsive $\Pi$ state is energetically 
close. Optical pumping of the system in the ground state thus results in two 
dissociation channels: one indirect route via the AC in the ground $\Sigma$ state 
and one direct path in the $\Pi$ state. We show that applying THz control pulses 
with specific time delays relative to the pumping, can significantly alter the 
population dynamics, as well as, the kinetic energy and angular distribution of 
the photofragments.
\end{abstract}

\maketitle

\section{Introduction}

Thanks to the continuously developing laser technology, which has
made it possible to generate light pulses with the length of few femtoseconds
or few hundred attoseconds, quantum control techniques are among the
most powerful tools of physics both in fundamental research and in
practical applications. The field of research is rapidly growing and
protocols have been adopted for studying different dynamical properties
and features of molecules starting from small diatomics to really
large polyatomic systems \cite{Rice1,Letokhov,Shapiro1,Regina1,Baumert1,Atabek1,Atabek2,Atabek3,Takatsuka1,Takatsuka2,Takatsuka3,Graham1,Tamar1,Tamar2,Tom1,Tom2,Tom3,Stefi1,Koch1,Koch2,Koch3,Sola1,Sola2,Sola3,Sola4,Sola5}. 

In recent years, efforts were invested to apply the dynamic Stark
effect (DSE) for control chemical dynamical processes \cite{Stolow1,Stolow2,Stolow3,Sussman1,Stolow4,Han1,Liu1,Mignolet_2019}.
It can be resonant or nonresonant depending upon the applied light
frequency. In the first situation the strong laser radiation fields
can couple any two electronic states of the molecule due to the electric
transition dipole moment and can also shape them. 
So-called light-induced nonadibatic
phenomena arise. Light-induced or ``dressed''
adiabatic potentials are formed, which incorporate the laser-molecule
coupling effects. Numerous theoretical and experimental studies have
demonstrated that the light-induced nonadiabatic phenomena (light-induced
avoided crossings or light-induced conical intersectons) have strong
impact on the dynamical and spectroscopic properties of molecular
systems \cite{Gabor1,Gabor2,Andris1,Andris2,Tamas1,Tamas2,Buksbaum1}. In the
second case, if the laser field is non-resonant with the energy difference
of any two electronic states of the molecule, still can have a significant
dynamical effect due to shaping of the potential energy surfaces through
the permanent dipole moments. This effect is very well studied in
the literature as it provides a general tool for quantum control of
atomic and molecular dynamical processes \cite{Stolow1,Stolow2,Stolow3,Sussman1,Stolow4,Han1,Liu1}.
The dynamic Stark effect can be described either in the
dipole or in the Raman limit. In the dipole limit the interaction
follows the instantaneous electric field, whereas in the Raman limit,
(when the dipole approximation is symmetry forbidden) the interaction
only follows the laser-pulse envelop \cite{Stolow4}. 

In the present work our showcase example is the lithium fluoride molecule,
therefore the control procedure relies on the dipole limit. The LiF
molecule has already been studied in our former works \cite{LiF_paper1,LiF_paper2}
where we discussed the role played by the lowest-lying $\Pi$
electronic state in the photodissociation of the molecule through
the population dynamics, the angular distribution and the kinetic
energy release (KER) spectra of the photofragments. Describing appropriately
the light-induced nonadiabatic phenomena the rotational degree of
freedom has already been taken into account as dynamical variable
in those works. Although in the present work we focus on different
subject and control the dynamics by a single cycle THz laser pulse,
the molecular rotation is also included in the numerical simulations
so as to describe accurately the photodissociation process. 

Recently, attention has been paid to control the dynamical and other
properties of molecules by single cycle THz pulses. Fleischer\emph{
et al}. investigated both theoretically and experimentally the THz-induced
molecular alignment in the gas phase using intense single-cycle THz
pulses \cite{Fleischer1}. This group has also studied experimentally
the decay of field-free rotational dynamics by terahertz-field-induced
molecular orientation \cite{Fleischer2}. Kurosaki \emph{et al}. proposed
a theoretical control scheme of temporal wavepacket separation for
oriented molecules. By using linearly polarized single-cycle THz pulse
they could separate the binary mixture of alkalihalide isotopologues
$^{133}$CsI and $^{135}$CsI \cite{Kurosaki}. Sub-one-cycle THz pulses 
were employed in the strategy suggested by Do\v{s}li\'{c} \cite{Doslic_2006} 
to achieve state-selective population transfer in the ACAC molecule.

In this article, we address another issue that is of similar importance.
Namely, the effect of a THz control pulse on the photodissociation 
process of a diatomic system. Our showcase example is the LiF molecule.

This paper is organized as follows: the working Hamiltonian and the
computational details of the calculations are explained in Sect. 2.
In Sect. 3, the results are presented and discussed. A summary and
conclusions are given in the final section.

\section{The Physical Situation and Methods}

\begin{figure}
 \includegraphics[clip,width=0.4\textwidth]{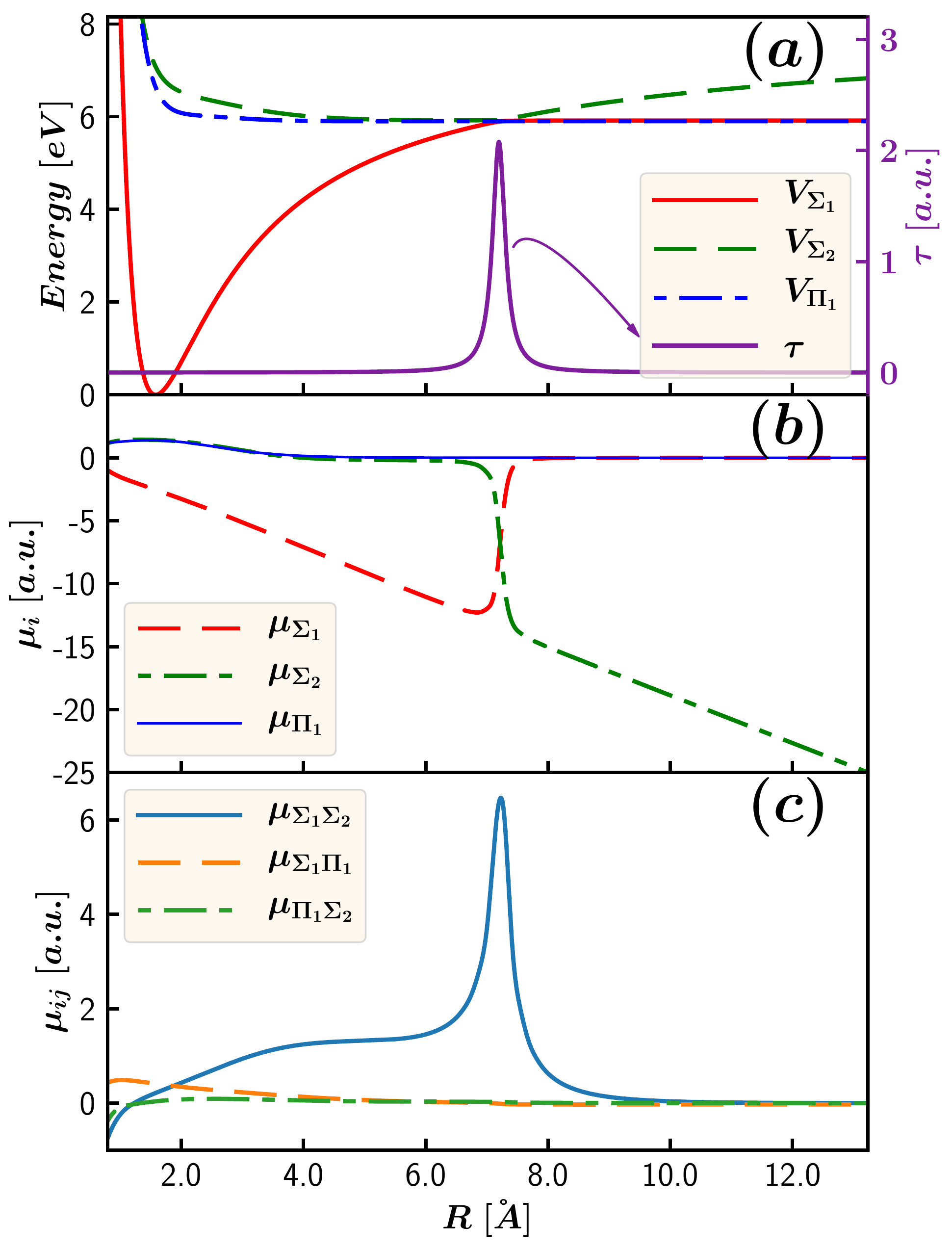}
 \caption{(Color online) (a) The lowest three adiabatic potential energy curves of the LiF molecule and the
nonadiabatic coupling term $\tau$(R) between the two $\Sigma$ states (scale is on the right side). 
(b) Permanent dipole moment functions of the three adiabatic electronic states.
(c) The transition dipole moment functions between the different electronic states.}
\label{Fig-1} 
\end{figure}
Lithim fluoride along the other alkali-halides have been a popular testing ground for nonadiabatic dynamics 
during the photodissociation of these molecules due to the avoided crossing (AC) between their lowest 
lying $^{1}\Sigma^{+}$ electronic states. In our previous works on this system we showed that a realistic 
theoretical description must also include the first $^{1}\Pi$ state \cite{LiF_paper1,LiF_paper2}. Accordingly, 
in the present investigation we model the LiF molecule as a three-level system considering the $1^{1}\Sigma^{+}$, 
$2^{1}\Sigma^{+}$ and $1^{1}\Pi$ electronic states, labeled throughout the paper as $\Sigma_1$, $\Sigma_2$ and 
$\Pi_1$. Their corresponding potential energy curves are presented in Fig.~\ref{Fig-1}(a), along the intrinsic 
nonadiabatic coupling term $\left(\tau (R) = \langle \varphi_{\Sigma_1} | \frac{\partial}{\partial R} \varphi_{\Sigma_2}\rangle\right)$ 
linking the $\Sigma_1$ and $\Sigma_2$ states at the AC around $R\sim7.2$ \AA. Panel b and c of Fig.~\ref{Fig-1} 
show the permanent $\left(\mu_{i}(R) = - \langle \varphi_i | \sum_k r_k | \varphi_i \rangle\right)$ 
and the transition dipole moments $\left(\mu_{ij}(R) = - \langle \varphi_i | \sum_k r_k | \varphi_j \rangle\right)$, 
respectively. An important feature of the transition dipole moments (TDM) is that the one responsible for 
the $\Sigma$-$\Sigma$ transitions, i.e. $\vec{\mu}_{\Sigma_1\Sigma_2}$, is parallel with the molecular axis
while the ones involving the $\Pi_1$ state are perpendicular.

Computation of the above electronic structure quantities of LiF have been 
carried out with the Molpro~\cite{molpro} program package at the 
MRCI/CAS(6/12)/aug-cc-pVQZ level of theory. In particular, the $\tau(R)$ has 
been computed by finite differences of the MRCI electronic wave functions.
The number of active electrons and molecular orbitals in the individual irreducible 
representations of the C$_{2v}$ point group were A$_1$ $\rightarrow$ 2/5, 
B$_1$ $\rightarrow$ 2/3, B$_2$ $\rightarrow$ 2/3, A$_2$ $\rightarrow$ 0/1.
With these parameters, we achieved a good agreement with the results 
of other studies~\cite{varandas1,pelaez,Triana_2019}.

\subsection{Working Hamiltonian}\label{Sec:Working-Ham}
As stated above, in our previous works on the LiF we showed that for a realistic 
description of the dynamics of the molecule one should consider all three electronic 
states ($\Sigma_1$, $\Pi_1$, $\Sigma_2$) in a theoretical calculation, and also its
rotational motion. Accordingly, the time-dependent Hamiltonian employed in the present 
investigation reads
\begin{widetext}
 \begin{equation}
  \hat{\mathrm{H}}=
  \begin{pmatrix}
   T & 0 & K \\
   0 & T & 0 \\
   K & 0 & T
  \end{pmatrix}
  +
  \begin{pmatrix}
   V_{\Sigma_1}-\mu_{\Sigma_1}\cos(\theta)E(t) & \quad -\mu_{\Sigma_1 \Pi_1}\sin(\theta)E(t)  & \quad -\mu_{\Sigma_1 \Sigma_2}\cos(\theta)E(t) \\
   -\mu_{\Sigma_1 \Pi_1}\sin(\theta)E(t)       & \quad V_{\Pi_1}-\mu_{\Pi_1}\cos(\theta)E(t)  & \quad -\mu_{\Pi_1 \Sigma2}\sin(\theta)E(t) \\ 
   -\mu_{\Sigma_1 \Sigma_2}\cos(\theta)E(t)    & \quad -\mu_{\Pi_1 \Sigma_2 }\sin(\theta)E(t) & \quad V_{\Sigma_1}-\mu_{\Sigma_2}\cos(\theta)E(t)
  \end{pmatrix}
  \label{H_sys}
 \end{equation}
\end{widetext}

Here, in the first term $T$ stands for the kinetic energy operator while $K$ is the 
intrinsic non-adiabatic coupling between states $\Sigma_1$ and $\Sigma_2$ at the 
avoided crossing. As we consider rotating-vibrating molecules, the kinetic energy term 
is given by
\begin{equation}
 T(R,\theta) = -\frac{1}{2 M_{r}} \frac{\partial^2}{\partial R^2} + \frac{L_{\theta}^2}{2 M_{r} R^2},
\end{equation}
where $R$ is the internuclear distance and $\theta$ is the angle between the laser polarization 
direction and the molecular axis, i.e. the the rotational coordinate. $M_r$ is the reduced mass, 
while $L_{\theta}$ is the angular momentum operator with $m=0$. For the nonadiabatic coupling 
operator we used an approximate form \cite{Coupling}
\begin{equation}
 K(R)\approx \frac{1}{2M_r}(2\tau(R)\frac{\partial}{\partial R}+\frac{\partial}{\partial R}\tau(R)),
\end{equation}
with $\tau$ being the nonadiabatic coupling term presented on Fig.~\ref{Fig-1}a.

\begin{figure}[ht]
\includegraphics[clip,width=0.45\textwidth]{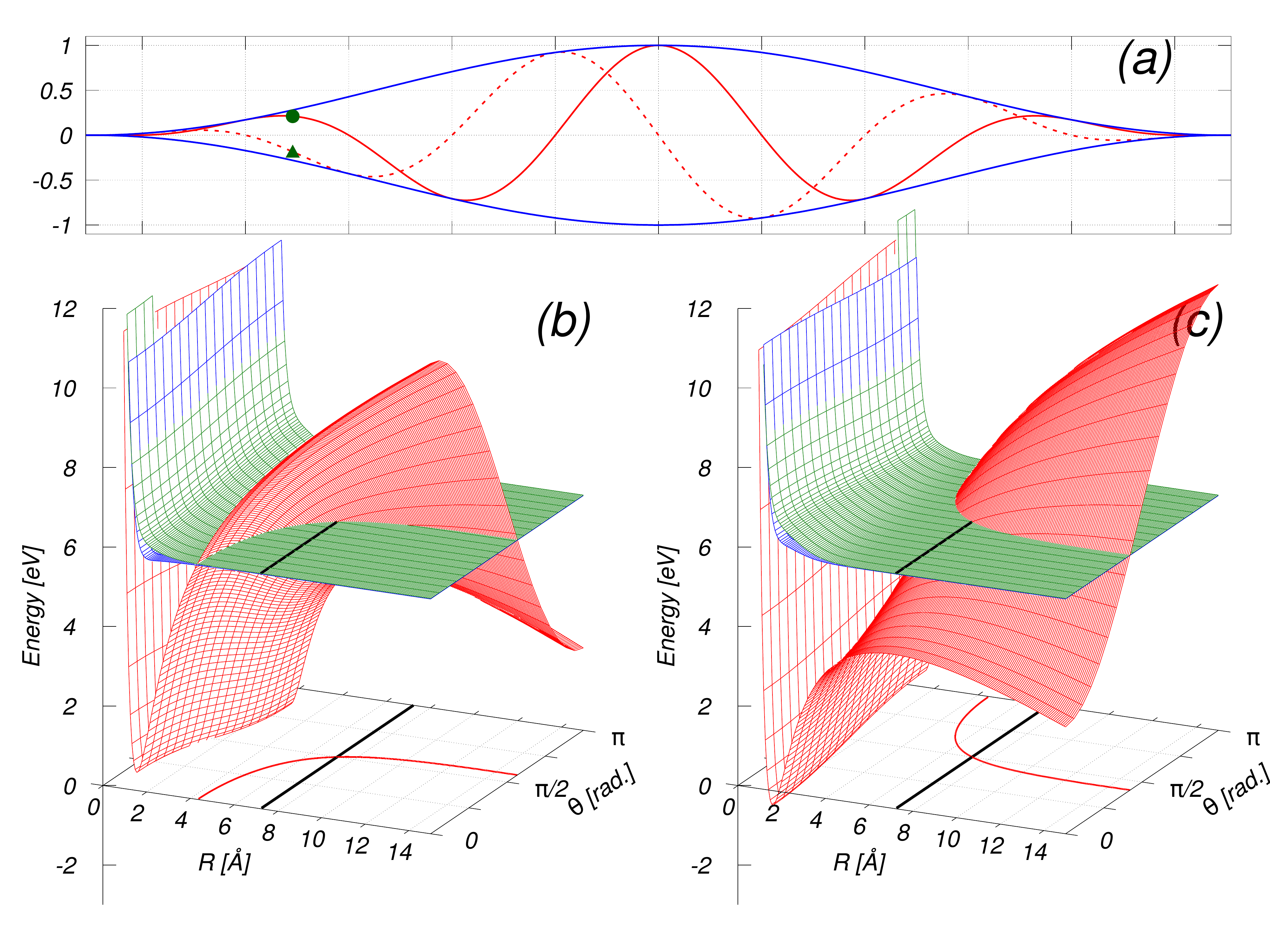} \caption{(a) General form of the THz electric 
fields applied in the present work ($\hbar\omega_c$ = 0.037 eV, I$_c =3.16\times10^{13} W/cm^2$). 
Two particular CEP cases of interest are presented ($\varphi_c=0$ with solid red line and $\varphi_c=\pi/2$ 
with dashed red line). The corresponding light induced potential energy surfaces (LIPs) are shown 
in panel (b) for $\varphi_c=0$ and in panel (c) for $\varphi_c=\pi/2$.}
\label{Fig-LIPs}
\end{figure}

The second term in the expression of $\hat{\mathrm{H}}$ is the potential energy matrix including 
the coupling with the applied $E(t)$ laser field. 
As the different potential energy surfaces are dipole coupled, we restrict this light-matter interaction 
to the first order DSE, i.e. the dipole limit.
Although the Hamiltonian of Eq. \ref{H_sys} was used throughout our calculations, it is easier to 
understand the system using the light induced potentials (LIPs), in terms of which the potential 
energy matrix is diagonal \cite{Scheit_JPCA2011}.
They are presented on Fig. \ref{Fig-LIPs}, and will served a pivotal role in the interpretation of our results.

Unless specified otherwise, atomic units with $e=m_e=\hbar=1$ are used throughout the article.

\subsection{The applied electric field}\label{Sec:E-Field}
In our calculations we used two linearly polarized (in the same direction) 
laser pulses, both of the form
\begin{equation}
 E(t,\varphi)=E_{0}f(t)\cos(\omega(t-t_{0})+\varphi),
 \label{Eq-laser}
\end{equation}
with cosine-squared envelopes
\begin{equation}
 f(t)=\cos^{2}\left( \frac{1.14372(t-t_{0})}{\tau}\right),
\end{equation}
where $\tau$ is the full width at half maximum (FWHM) of the intensity 
profile. The dynamics was initiated by a $\tau_{p}=20$ fs long pump pulse, 
which also defined the origin of our time axis, that is $t_{0p}=0$. For 
all the results presented in this work, the energy of the pump was fixed 
to $\omega_p=6.94$ eV, and its intensity to $I_p=5\times10^{12}\mathrm{W/cm^2}$.

The second one was a single cycle THz pulse with $\omega_c=0.037$ eV, with 
the corresponding pulse duration $\tau_c=111.77$ fs, and $I_c=3.16\times10^{13}\mathrm{W/cm^2}$. 
This control field is unable to produce transitions between the electronic states, 
however it alters the potential energy landscape of the molecule, which has
a great impact on the outcome of the photodissociation process. Two ``control knobs'' 
were chosen to steer the systems dynamics: the $\varphi_c$ carrier envelope phase (CEP) 
of the control pulse, and the time delay $\Delta t=t_{0c}-t_{0p}$ between the pulses.

\subsection{Propagation of the wave packets}\label{Sec:Propag-WP}
The time-dependent Schr\"{o}dinger equation (TDSE) that described the dynamics of
the system was solved using the 
MCTDH (multi configurational time-dependent Hartree) method~\cite{MCTDH1,MCTDH2,mctdh_4,*mctdh_4_1}. 
The vibrational degree of 
freedom ($R$) was described by a sin-DVR primitive basis with $N_{R}$ basis 
elements distributed between 0.79 and 31.75 \AA\quad for the internuclear separation.
For the description of the rotational degree of freedom ($\theta$) Legendre 
polynomials $\left\{ P_{j}(\cos\theta)\right\} _{j=0,1,2,\cdots,N_{\theta}}$ 
were used. These primitive basis sets ($\chi$) were employed to represent the 
single particle functions ($\phi$), which in turn were used to build up the nuclear 
wave function ($\psi$):
\begin{align}
 \phi_{j_{q}}^{(q)}(q,t) &= \sum_{l=1}^{N_{q}}c_{j_{q}l}^{(q)}(t)\;\chi_{l}^{(q)}(q)\qquad , q=R,\,\theta \nonumber\\
 \psi(R,\theta,t) &= \sum_{j_{R}=1}^{n_{R}}\sum_{j_{\theta}=1}^{n_{\theta}}A_{j_{R},j_{\theta}}(t)\phi_{j_{R}}^{(R)}(R,t)\phi_{j_{\theta}}^{(\theta)}(\theta,t). 
 \label{psi_MCTDH}
\end{align}
In our numerical calculations $N_{R}=2048$ and $N_{\theta}=361$ primitive basis 
functions were used. In order to ensure the correct convergence of the propagations, 
on all adiabatic surfaces and for both degrees of freedom a set of $n_{R}=n_{\theta}=50$ 
single particle functions were used to build up the nuclear wave function of the system. 
This relatively high value was necessary  as the THz control field induced a considerable 
amount of rotation.

\begin{figure*}[ht!]
 \begin{subfigure}[t]{0.49\textwidth}
  \centering
  \includegraphics[width=0.95\textwidth]{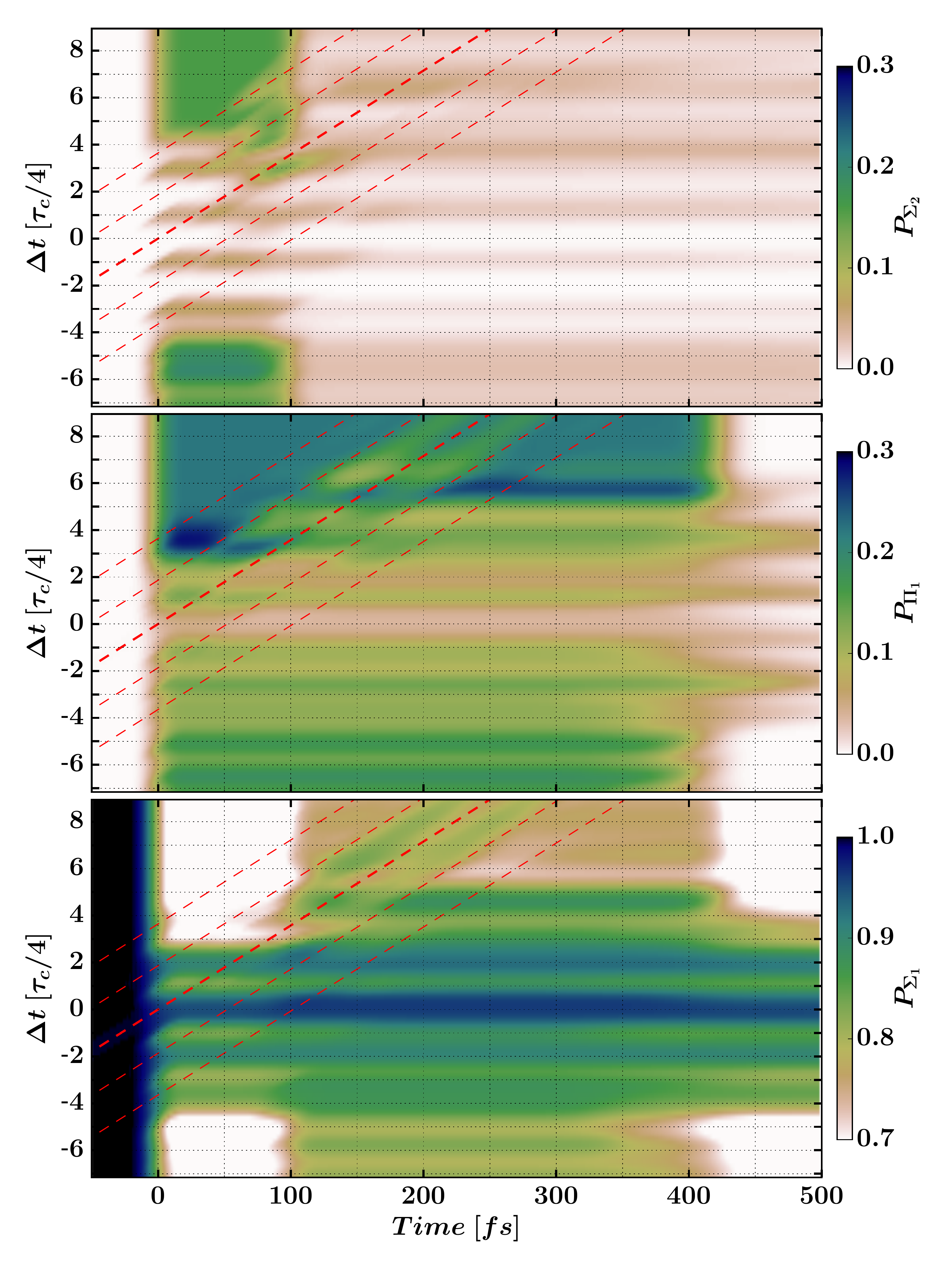}
  \caption{$\Delta t$ scan for $\varphi_c$=0.}
 \end{subfigure}
 \hfill
 \begin{subfigure}[t]{0.49\textwidth}
  \centering
  \includegraphics[width=0.95\textwidth]{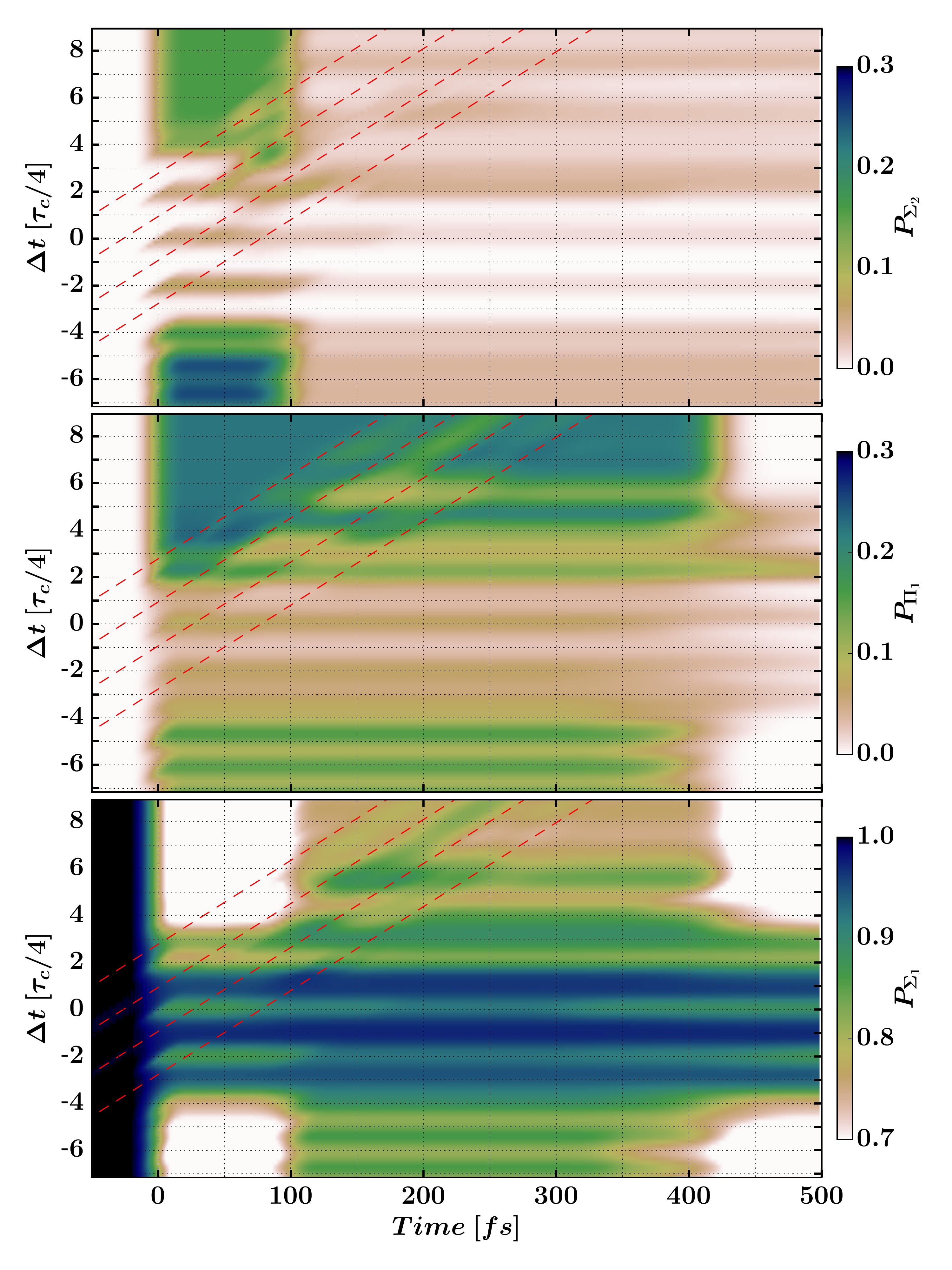}
  \caption{$\Delta t$ scan for $\varphi_c$=$\pi/2$}
 \end{subfigure}
 \caption{Evolution of the state population as a function of $\Delta t$ for two different carrier-envelope phases. 
 Transverse red lines mark the time moments when the electric field of the control pulse has minima/maxima.}
 \label{Fig_StatePop}
\end{figure*}

\subsection{Calculated quantities}\label{Sec:Quantities}
The solutions of the TDSE were then used to calculate the populations of the employed 
electronic states~\cite{LiF_paper1}, the kinetic energy release spectra (KER) and the 
angular distribution of the molecular fragments~\cite{Gabor5}. The electronic state 
populations are obtained as:
\begin{align}
 P_{i}(t)&=\left\langle \psi_{i}(R,\theta,t) \left|\right. \psi_{i}(R,\theta,t) \right\rangle \quad\quad,i\in\left\{ \Sigma_1,\Pi_1,\Sigma_2\right\} \nonumber\\
 &=\int_{0}^{\pi}\mathrm{d}\theta\cdot\sin\theta\int_{0}^{\infty}\mathrm{d}R\cdot\psi_{i}^{*}(R,\theta,t)\cdot\psi_{i}(R,\theta,t),
\end{align}
where $\psi_i$ are the projections of the total nuclear wave function of Eq. \ref{psi_MCTDH} on the considered electronic states.
The KER is calculated according to the following formula:
\begin{equation}
 P_{KER}^{i}(E)=\int_{0}^{\infty}\mathrm{d}t~\int_{0}^{\infty}\mathrm{d}t' \left\langle \psi_{i}(t)\left| W \right| \psi_{i}(t')\right\rangle \mathrm{e}^{-\mathrm{i} E(t-t')},
\end{equation}
where $-\mathrm{i}W$ is the complex absorbing potential (CAP) applied at the 
last 5.29 \AA\, of the grid related to the vibrational degree of freedom 
of each electronic state.
The angular distribution of the photofragments is given by:
\begin{equation}
 P_{ang}^{i}(\theta_{j})=\frac{1}{w_{j}}\int_{0}^{\infty}\mathrm{d}t~\left\langle \psi_{i}(t)\left| W_{\theta_{j}} \right| \psi_{i}(t)\right\rangle,
\end{equation}
where $-\mathrm{i}W_{\theta_{j}}$ is the projection of the CAP to a specific 
direction of the angular grid ($j=0,\ldots,N_{\theta}$), and $w_{j}$ is the DVR 
weight associated to this grid point. In the last two equations the superscript $i$ 
stands for either $\Sigma_1$ or $\Pi_1$ as the molecule can dissociate on these two states.

\section{Results and discussion}

The dynamic Stark effect is usually examined as a function of the time delay between the pump and the Stark (control) pulse. We follow this tradition and 
start our investigations by looking at the evolution of the state populations changing the center of the control pulse from t$_{0c}$=-200 fs to 255 fs. The results are 
presented on Fig. \ref{Fig_StatePop} (a) and (b) for $\varphi_c=0$ and $\varphi_c=\pi/2$, respectively. Here, the time delay is conveniently expressed in units of the 
control pulse period, $\tau_c=111.77$ fs. Also, to help understand the data, red transverse dashed lines mark the time moments when the control 
field has an extrema (minima or maxima). The spacing between these lines is not $\tau_c/2$ as the envelope of the pulse ``pushes'' 
the field extrema slightly toward the center of the pulse. From this figure it is clear that the choice of $\Delta t$ has a huge impact on the behavior of 
the system. This behavior differs however in a few key aspects from that found in the literature of the dynamic Stark effect. Those works almost exclusively describe 
the nonresonant dynamic Stark effect (NRDSE) in the moderately intensive (non-perturbative but non-ionizing) regime and the Raman limit. As a consequence of the Stark shifted potentials 
the velocity with which the excited wavepacket traverses the crossing region is altered, and according to the Landau-Zener formula \cite{landau_zener} the branching ratio 
of the photofragments is modified. This control scheme is most pronounced when the Stark field is applied either during the pump process or when the wavepacket is around 
the crossing point. If it comes before or after these time moments, the dynamics of the system remains unaffected.

The fundamental difference in the present work, as mentioned above, is that the electronic states are dipole coupled meaning that the first order DSE applies, hence the 
interaction follows the instantaneous electric field. Besides, the intensity of our control pulse, while still non-ionizing, is relatively high, which combined with the 
first order DSE leads to significant modifications of the potential surfaces, as illustrated by Fig \ref{Fig-LIPs}. This leaves pronounced changes in the evolution of the 
state populations presented on Fig. \ref{Fig_StatePop}, for all investigated time delays (in each case the nuclear time-dependent Schr\"{o}dinger equation was propagated 
beyond 1 ps, but most of the dynamics ceased around 400 fs, when the dissociating wavepackets reach the absorbing potential at the end of the numerical grid). The most 
striking feature is the suppression of the pump process when the two pulses overlap. In this interval the excited populations are not only decreased, but also show a 
modulation as a function of delay time, which resembles the periodicity of the control pulse. Interestingly, similar modulations are present even when the control pulse 
precedes the pump. The other important phenomena that has to be noted is that after the dynamics is initiated there is usually a population transfer around the control 
field extrema, which in turn impacts the branching ratio between the dissociation channels $\mathrm{LiF} \rightarrow \mathrm{Li} + \mathrm{F}(^2P_{1/2})$ and 
$\mathrm{LiF} \rightarrow \mathrm{Li} + \mathrm{F}(^2P_{3/2})$ correlating to the $\Pi_1$ and $\Sigma_1$ states, respectively.

\begin{figure}[ht!]
 \includegraphics[width=0.5\textwidth]{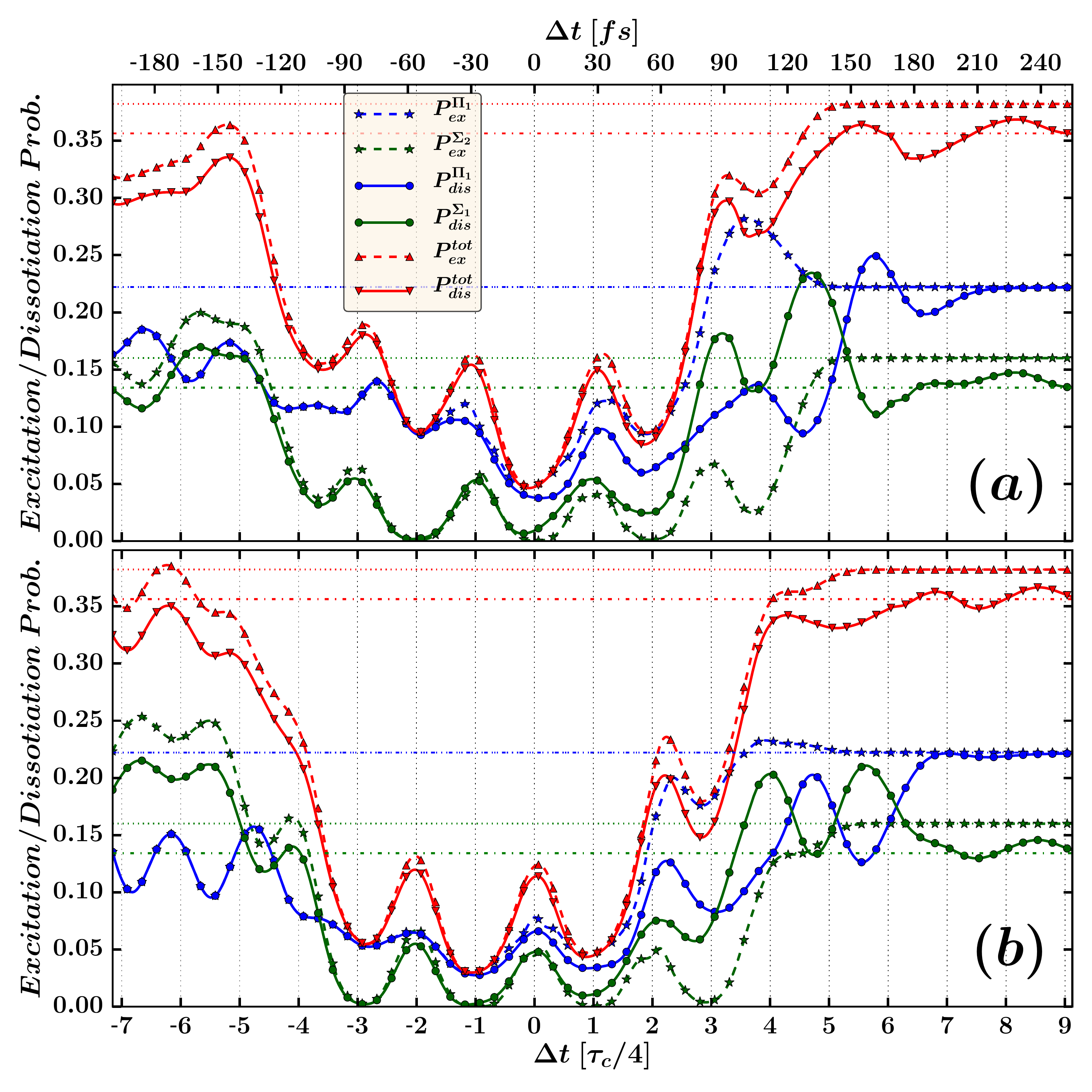}
 \caption{Excited population on the $\Pi_1$ and $\Sigma_2$ states marked with blue and green dashed lines, 
 and dissociation probability on the $\Pi_1$ and $\Sigma_1$ states marked with blue and green continuous lines. 
 Red lines represent the sum of the excited and dissociated population of the above two channels. The control 
 pulse carrier-envelop phase is: (a) $\varphi_c=0$, (b) $\varphi_c=\pi/2$.}
 \label{Fig_ExDiss}
\end{figure}

In order to better visualize the above findings, we present on Fig. \ref{Fig_ExDiss} the excitation (dashed lines with stars) and dissociation (full curves with circles) 
probabilities of the different channels. Green and blue lines stand for quantities related to the $\Sigma_2$ (dissociation in $\Sigma_1$) and the $\Pi_1$ states, respectively, 
while red curves represent their sum. Also, horizontal dotted and dashed-dotted lines with the same color-coding mark the excitation and dissociation probability of the system 
in the absence of the THz pulse. As $\Pi_1$ is a fully dissociative state the related horizontal blue lines overlap. It can be seen, that after sufficiently long delay times 
($\sim\,5\tau_c/4$) the populations pumped to the excited states converge to their values obtained in the control-free case. As mentioned above, during the overlap of the 
pump and the control fields the excitation efficiency is greatly reduced, and takes place in short bursts around the time moments when the instantaneous control field is zero. 
This is more pronounced for the $\Sigma_2$ state, which is practically unaffected by the pump when the electric field of the control pulse has an extrema. It is worth mentioning, 
that in this $\Delta t$ range almost none of the excited population remains trapped on the $\Sigma_2$ state, as indicated by the proximity of the total excitation and dissociation 
curves. For smaller delay times, when the control pulse terminates before the pump is switched on, the excited population on the $\Pi_1$ state remains below its control-free 
value, while the one on the $\Sigma_2$ exceeds it. This is more prominent for the $\varphi_c=\pi/2$ case.

\begin{figure*}[t!]
 \begin{subfigure}[t]{0.49\textwidth}
  \centering
  \includegraphics[width=0.95\textwidth]{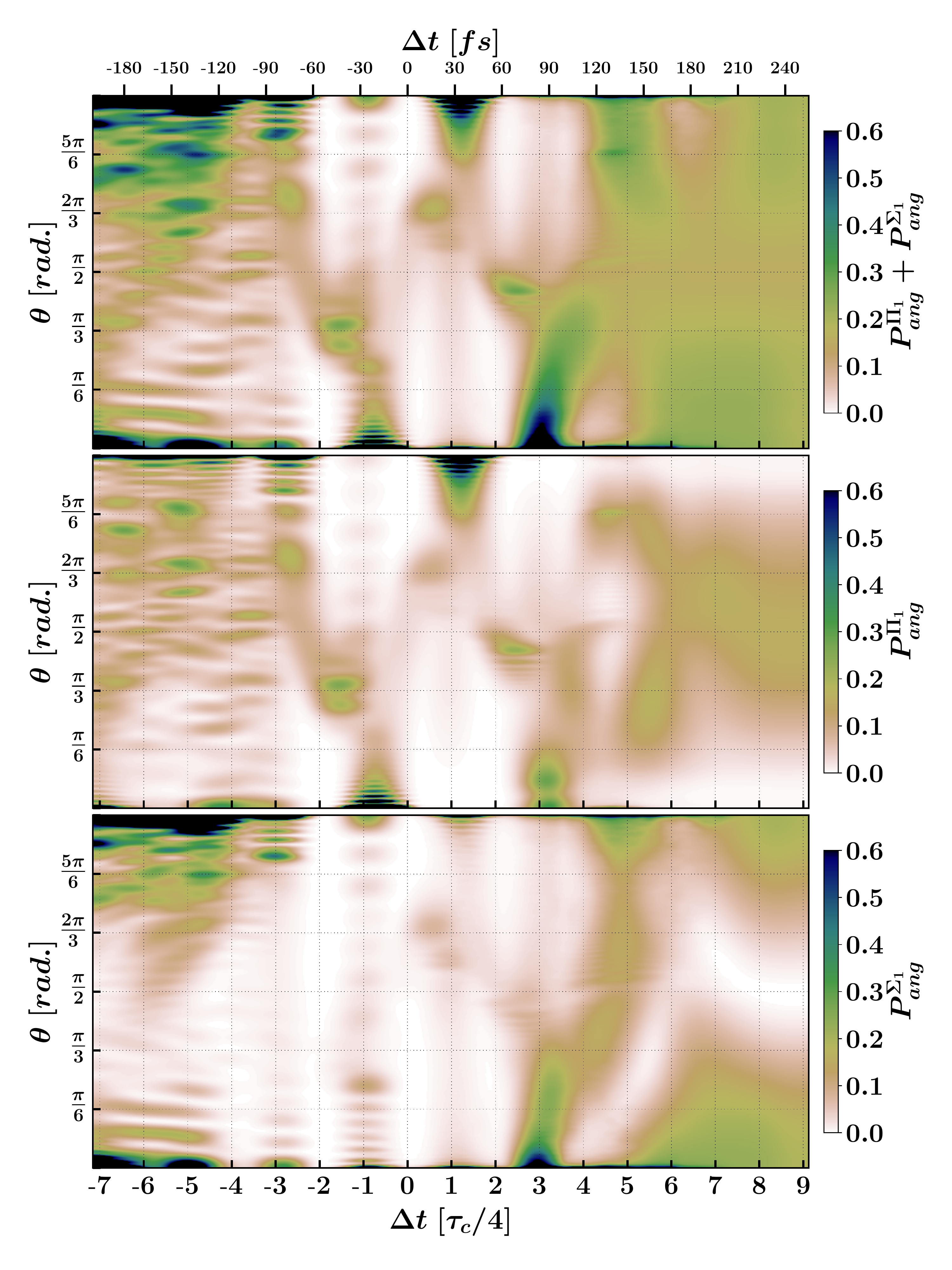}
  \caption{$\Delta t$ scan for $\varphi_c$=0.}
 \end{subfigure}
 \hfill
 \begin{subfigure}[t]{0.49\textwidth}
  \centering
  \includegraphics[width=0.95\textwidth]{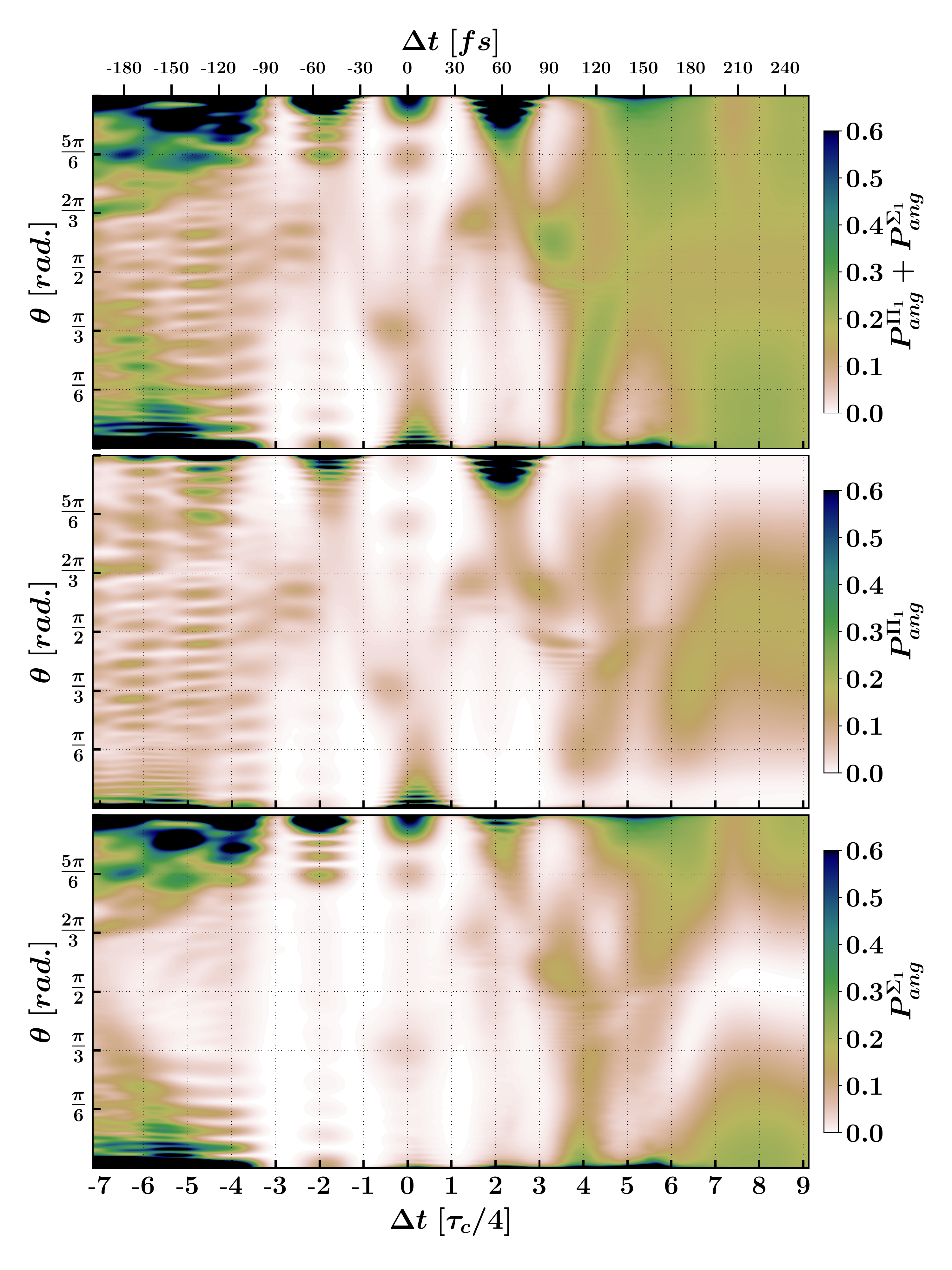}
  \caption{$\Delta t$ scan for $\varphi_c$=$\pi/2$}
 \end{subfigure}
 \caption{Angular distribution of the photofragments on the $\Sigma_1$ (bottom) and $\Pi_1$ (middle) electronic states, 
 and the sum of the two channels (top) as a function of time delay between the pump and the control pulses.}\label{Fig_AngDist}
\end{figure*}

The above detailed behavior of the pump process is rooted in our theoretical description, that is the consideration of the rotational degree of freedom. Traditional NRDSE 
control techniques rely on the use of infrared pulses, which are unable to produce any transitions in the investigated system (hence the name non-resonant). In contrast, we chose 
to work with a THz radiation, which albeit is still unable to produce electronic transitions, it induces rotational and vibrational excitations. These rotational excitations 
persist even after the control pulse is finished. This leads to the oscillation of a rotational wave packet on the ground electronic state, which is the cause of the modulations 
observed in the populations excited to the $\Sigma_2$ and $\Pi_1$ states for large negative delay times, as the pump pulse no longer encounters the original isotropic initial distribution. 
Moreover, the interference between the various components of this wave packet leads to the development of a nodal structure, which also manifests in the angular distribution of the 
photofragments. These angular distributions for the two considered carrier-envelope phases of the control pulse, $\varphi_c=0$ (left panel) and $\varphi_c=\pi/2$ (right panel), 
are presented on Fig. \ref{Fig_AngDist} for the two distinct dissociation channels and also their sum. The figure shows, that the dissociation occurs primarily along the (common) 
polarization axis of the employed laser pulses, and as Fig. \ref{Fig_ExDiss} already suggested, mostly on the $\Sigma_1$ state. Considering the nature of the transition dipoles 
($\mu_{\Sigma_1\Sigma_2}$/$\mu_{\Sigma_1\Pi_1}$ is parallel/perpendicular to the molecular axis) this means that the THz pulse oriented the molecules along its polarization axis, 
and this orientation remained, or more precisely it was periodically partially revived, after the pulse ended.

The suppression of the pump process during the temporal overlap of the two laser pulses can be best understood based on 
the light induced potentials presented on Fig. \ref{Fig-LIPs}. In order to have an efficient population transfer between 
two dipole coupled electronic surface, two conditions have to be met: the coupling radiation has to be resonant for a given 
region (($R$, $\theta$) in our 2D case) of the involved surfaces, and these regions need to be populated. As we saw earlier, 
the THz control pulse induces a rotational excitation of the system. Moreover, as the LiF is a polar molecule, the control 
field orients the molecule instead of aligning it. In the LIPs picture this manifests in the deformation of the potential 
surfaces along the $\theta$ coordinate: for a given internuclear distance, the PES ascends or descends compared to its 
field free position along the $\theta$ direction due to the $\mu_i\cos(\theta)E(t)$ term of the Hamiltonian. In other words, 
a potential well forms around $\theta=\{0, \pi\}$ periodically. For our initial isotropic distribution in the ground state 
this means a periodic concentration in these potential wells, i.e. up/down orientation of the molecules. Another important 
factor is that the permanent dipoles of the excited states have opposite signs compared to the ground state PDM in the Franck-Condon 
region, which means that they are displaced in the opposite direction than $\Sigma_1$. 
Accordingly, when the molecules are oriented either up or down, the detuning between the $\Sigma$ states exceeds the pump energy
and instead of an increased excited population we end up with none. The condition of population transfer to $\Sigma_2$
exist only in a short time window around the time moments when the control field is zero which again reduces the pump efficiency.
\begin{figure*}[t!]
 \begin{subfigure}[t]{0.49\textwidth}
  \centering
  \includegraphics[width=0.95\textwidth]{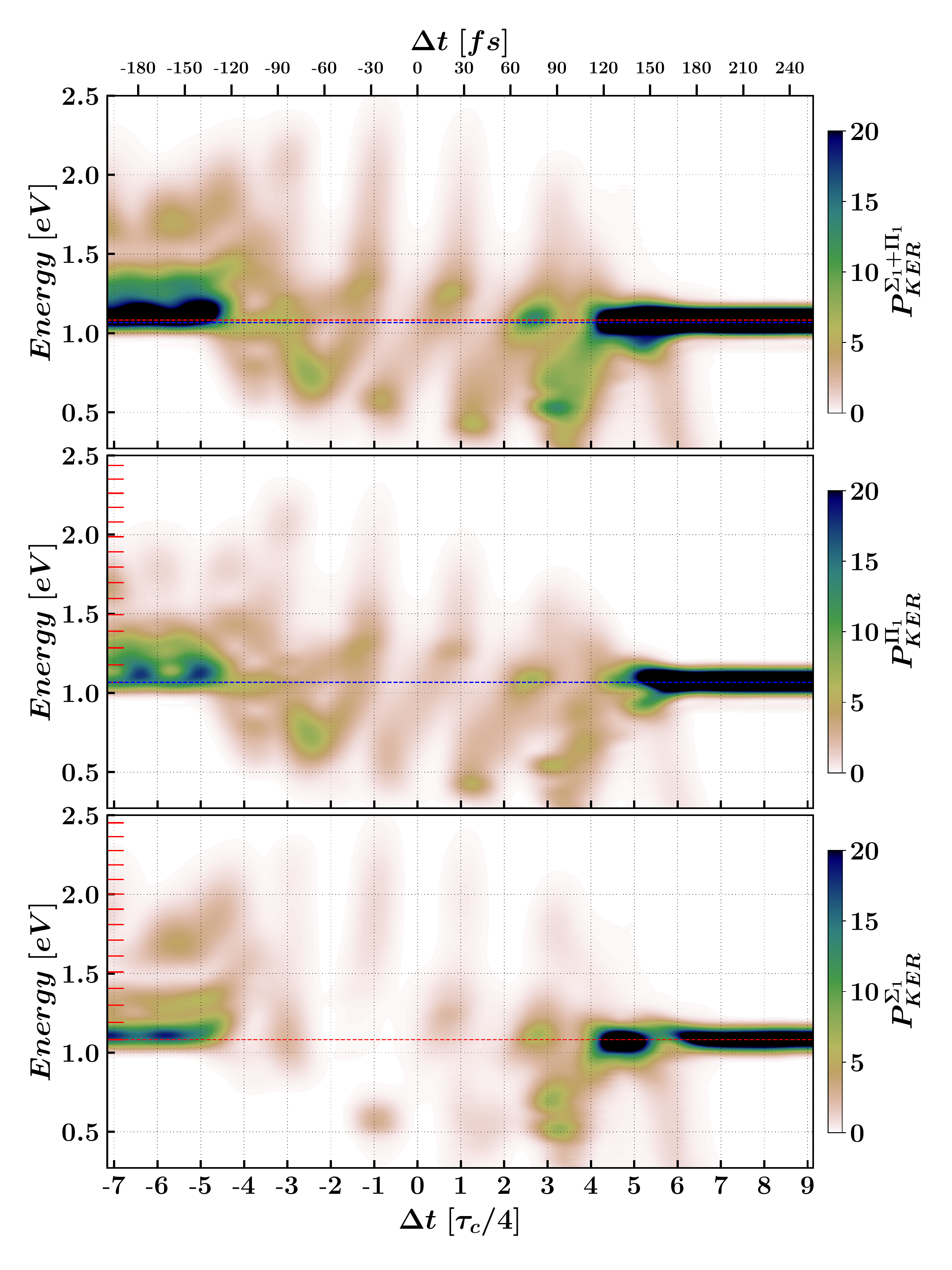}
  \caption{$\Delta t$ scan for $\varphi_c$=0.}
 \end{subfigure}
 \hfill
 \begin{subfigure}[t]{0.49\textwidth}
  \centering
  \includegraphics[width=0.95\textwidth]{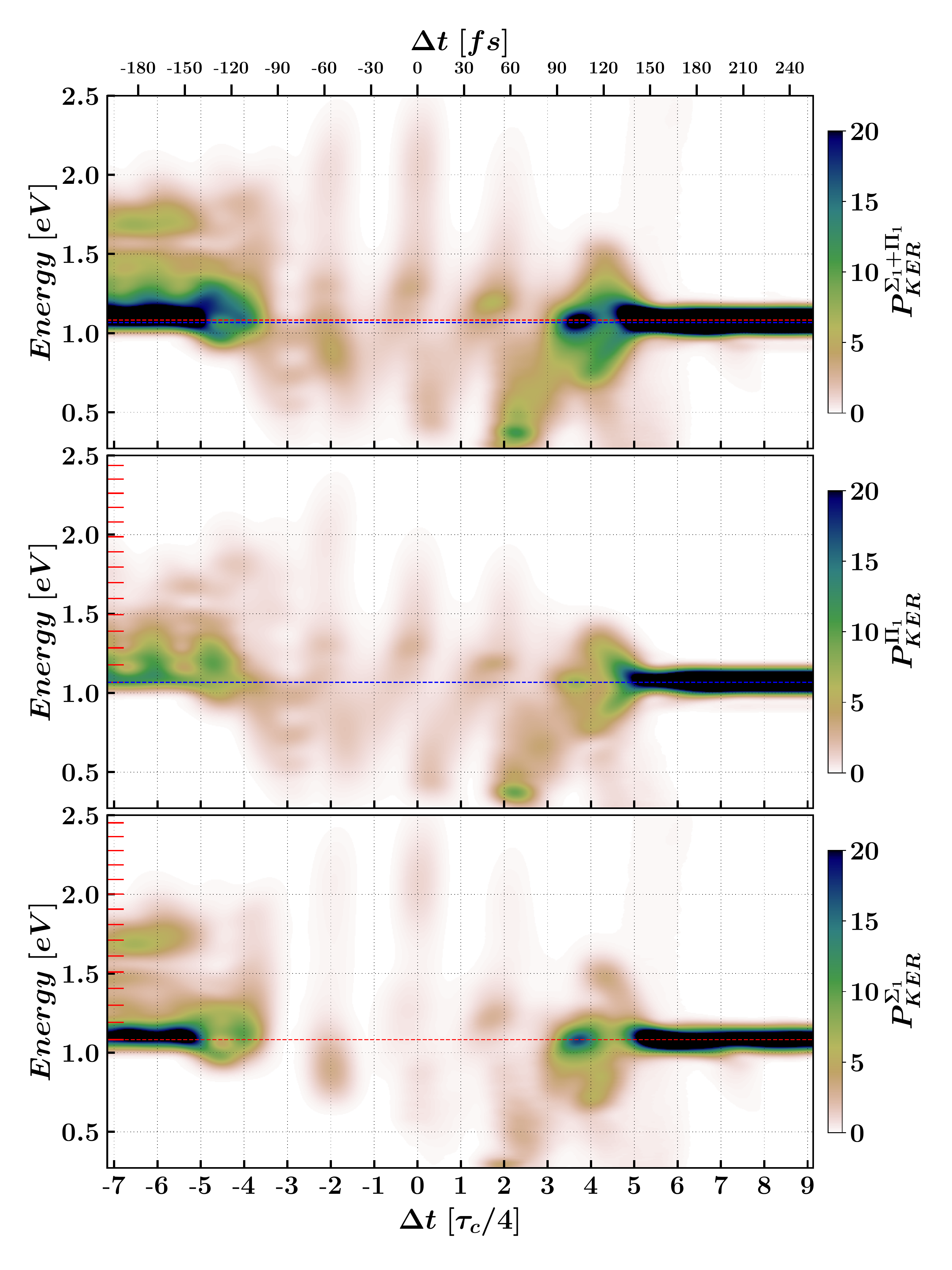}
  \caption{$\Delta t$ scan for $\varphi_c$=$\pi/2$}
 \end{subfigure}
 \caption{Kinetic energy release distribution of the photofragments on the $\Sigma_1$ (bottom) and $\Pi_1$ (middle) electronic states, 
 and the sum of the two channels (top) as a function of time delay between the pump and the control pulses.}\label{Fig_KER}
\end{figure*}

The situation of the $\Pi_1$ state is more interesting. The fundamental difference here is that the TDM with the ground state is 
perpendicular to the molecular axis\cite{LiF_paper1}. This means that the two states are coupled in the region where the control field least 
distorts the potential surfaces (see again Fig. \ref{Fig-LIPs}). Due to these facts, intuitively one would expect most of the 
dissociating fragments to be detected perpendicular to the laser polarization direction, however this is not the case. Having 
in mind that the pump energy was tuned to the $\Sigma_1$-$\Sigma_2$ transition, it is easy to see that the resonance condition 
between $\Sigma_1$ and $\Pi_1$ is shifting along the $\theta$ coordinate as the PESs swing under the action of the control pulse. 
This movement of the resonance point can be identified in the angular distributions $P_{ang}^{\Pi_1}$ in the $\Delta t\in[-3\tau_c,5\tau_c]$
delay time interval, although it is not a one-to-one correspondence, as the excited wavepacket is slightly (due to the considerably 
smaller $\mu_{\pi_1}$ than $\mu_{\Sigma_1}$) rotated on the distorted PES. More surprising is that in this interval the molecules 
dissociate with highest probability along the laser polarization direction. This can be understood in light of the wavepacket 
dynamics on the LIPs described earlier. As we saw, the control pulse orients the molecules up or down. Due to the fact, that 
$\Pi_1$ lays lower in energy than $\Sigma_2$, the resonance condition with the ground state along the polarization axis is 
achieved before the control field changes its sign (hence, the peaks are shifted from the zero control field moments). 
Accordingly, most of the ground state population is still concentrated in this (up/down) region, which results in a higher 
transition probability to $\Pi_1$ despite the reduced coupling. Moreover, as the field changes sign, the excited states develop 
potential wells in the direction where previously $\Sigma_1$ had (up/down), which results in the rotation of the $\Pi_1$ wavepacket 
toward the pump-forbidden $\theta=\{0,\pi\}$ direction. This in turn leads to the development of the interference structures
observable in the angular distribution \cite{LiF_paper2}. If the control pulse is applied after the system is pumped but before 
the excited wavepackets reach the AC region the angular distributions are more structured owing to the previously mentioned 
population transfer between the various states. This is most evident by the appearance of dissociating fragments around the 
perpendicular direction on $\Sigma_1$ accompanied by a reduced dissociation probability at the same time delays on $\Pi_1$. 
Finally, if the control pulse is turned on after the excited wavepacket traverses the AC region, the angular distributions converge 
to their usual control-free dipole shapes.

The Stark deformation of the potential surfaces depend on a number of factors: control field intensity, 
internuclear distance dependence of the permanent dipoles and orientation of the molecules. In addition, the used control field 
changes sign a number of times, which leads to an intricate wavepacket dynamics. Following this dynamics for each considered time 
delay is a cumbersome task which extends beyond the purpose of the present work. However, the main mechanisms shaping the response of 
the system toward the interaction with the control field can be identified. 

The effect of the control pulse on the excitation process was detailed above based on the angular distribution of the photofragments. 
A complementary information is provided by the kinetic energy release spectra of the dissociation products. These are presented on 
Fig. \ref{Fig_KER} in a similar arrangement as the angular distributions of Fig. \ref{Fig_AngDist}. Red and blue horizontal dashed lines 
mark the center of the KER spectra (Lorentzian shaped due to the single-photon pump process) in a pump-only scenario on the $\Sigma_1$ 
and $\Pi_1$ states, respectively. These results consolidate what we observed earlier. For large negative time delays we see that 
higher energies are present in the spectra, indicating that the molecules were ro-vibrationally excited in the ground state before 
the pump induced transitions to the excited electronic states. In the other extreme, for large positive delays, just as the angular 
distributions, the KER spectra also converge to their control-free value. In-between, when the control pulse is present while the 
excited wavepackets reach the AC, the spectra are smeared both to higher and lower energies than in the control-free case. This is 
the result of two processes. 

First of all, as the PESs are fluctuating under the action of the control pulse, the potential energy 
of the dissociating wavepackets is altered, which ultimately translates to modifications of the final kinetic energy of the photofragments. 
Whether it is increased or decreased depends on which region ($\theta<\pi/2$ or $\theta>\pi/2$) of the excited surface was the population 
placed on, and the phase of the control pulse (direction of the electric field). The magnitude of the energy shift follows the 
$\theta$-dependence of the PES modulations (strongest for the direction parallel with the laser field and non in the perpendicular direction).
This is reflected in the fact that the smallest KER values are obtained whenever the population is pumped in the direction of 
the field, where as we saw while discussing the angular distributions, the excited surfaces develop potential wells.

The second process is the above mentioned population transfers observed in the $\sim\Delta t\in[0,7\tau_c/4]$ interval.
This can also be attributed to the dynamically changing potential surfaces. Earlier works found in the literature \cite{Scheit_JPCA2011,Scheit_JCP2014} 
pointed out that in a diabatic picture the dynamically Stark shifted potentials also imply that the position of the crossing between 
the non-adiabatically coupled $\Sigma_1$ and $\Sigma_2$ states of LiF also changes as a function of time. This is illustrated on 
Fig. \ref{Fig-LIPs}, where the continuous black lines in the ($R$-$\theta$) plane mark the position of the intrinsic AC in the control-free 
case, while the red curve indicates the crossing between the light induced potentials at a given time moment during the action of 
the control pulse, marked by a green circle and triangle on the plot of the electric fields of Fig. \ref{Fig-LIPs} a). Moreover, 
in our three-state description a new light induce crossing emerges between the $\Sigma_1$ and $\Pi_1$ states (due to the proximity 
of $\Sigma_2$ and $\Pi_1$ this crossing is close to the $\Sigma_1$-$\Sigma_2$ one, and for clarity, only the latter one is plotted 
on the figures). It is obvious, that the instantaneous field intensity determines how much the dynamical crossing is shifted 
from its field-free position. Furthermore, we pointed out above that the LIPs swing along the $\theta$ coordinate, which leads to 
the $\theta$-dependence of the light-induced crossings. Whenever the dissociating wavepacket encounters these dynamically shifting 
LIP intersections it bifurcates, leading to population transfers between the involved surfaces. This takes place when the crossing 
is shifted to smaller internuclear distances, i.e. where $\Sigma_1$ is lifted upwards. As a result the population transfered to this 
state encounters a potential barrier and looses some of its kinetic energy before being transfered back to the excited states during 
the descending edge of the control pulse peak, when the crossing moves from smaller to larger internuclear distances. 

However, it is hard to distinguish the above two effects in the KER spectra, this later one is more prominent in the state populations 
of Fig. \ref{Fig_StatePop}. Here, for positive time delays the control field is strong enough to shift the crossings in the path of 
the dissociating wavepackets. As the $\Pi_1$ PES lies lower in energy, it is encountered first by the ascending $\Sigma_1$ surface, 
and part of the population is transfered. Immediately afterward the $\Sigma_1$-$\Sigma_2$ bifurcation occurs, whereupon part of the 
population initially pumped to the $\Pi_1$ state gets on $\Sigma_1$. On the descending edge of the pulse peak the situation is reversed, 
and due to the stronger $\mu_{\Sigma_1\Sigma_2}$ TDM most of the dissociating population on $\Sigma_1$ is transfered to $\Sigma_2$ and 
only a small amount returns to $\Pi_1$. Accordingly, the control pulse unidirectionally modifies the branching ration of the dissociation 
products favoring the $\Sigma_1$ channel. This effect seems to be the strongest around the highest central control field peak for both 
investigated CEP values, however this is somewhat hard to assess, as in this delay time region the initial excited populations are not 
the same due to the overlap of the two pulses.

Finally, the control pulse modifies not only the branching ratio of the dissociation channels, but also alters the amount of population temporally 
trapped on the bound $\Sigma_2$ state. This effect is best observed in the modulation of the difference between the total excitation and dissociation 
probabilities of Fig. \ref{Fig_ExDiss} (red curves). This happens for larger delay times, when the trailing edge of the control pulse Stark shifts the 
the potential surfaces only when the dissociating wavepackets are already in the neighborhood of the AC leading to the well studied modulations \cite{Stolow3,Stolow4,Graham1}
of the population transfer between the non-adiabatically coupled surfaces. 

\section{Conclusions}

In this work we have investigated the effect of a THz control pulse on the photodissociation process of the LiF molecule. Beside the vibrational degree of freedom 
our description also incorporated the rotational motion of the molecule. For the employed control frequency we saw that this choice is indispensable for a realistic 
description of the systems dynamics, as it greatly impacted the pump efficiency and the direction of the dissociating fragments. Also, the control pulse induced Stark 
fluctuations of the potential surfaces led to modulations of the kinetic energy release spectra and the appearance of new dynamically shifting surface crossings. 
As the dissociating wavepackets encountered these crossings population transfers occurred, which led to a modulation of the $\Sigma_1/\Pi_1$ dissociation branching 
ratio in favor of the former. Changing the carrier-envelope phase of the control pulse altered the timing of the population transfers, but otherwise did not impact
the outcome of the dissociation process significantly.

\section*{Acknowledgment}

This research was supported by the EU-funded Hungarian grant EFOP-3.6.2-16-2017-00005, and the ELI-ALPS project GINOP 2.3.6-15-2015-00001.
We are grateful to NKFIH for support (grant no. K128396). The supercomputing service of NIIF has been used for this work.

\bibliography{LiF_DSC-THz}

\end{document}